# Studies of Reynolds Stress and the Turbulent Generation of Edge Poloidal Flows on the HL-2A Tokamak

Ting Long[1], Patrick Diamond[1,2], Min Xu[1], Rui Ke[1], Lin Nie[1], Dong Guo[3] and HL-2A Team

[1] Southwestern Institute of Physics, Chengdu, China

[2] CASS and Dept. of Physics, University of California, San Diego, California, USA

[3] ENN Science and Technology Development Co., Ltd., Langfang, China

E-mail contact of main author: longt@swip.ac.cn

## Abstract

Several new results in the physics of edge poloidal flows, turbulent stresses and momentum transport are reported. These are based on experiments on the HL-2A tokamak. Significant deviation from neoclassical prediction for mean poloidal flow in Ohmic and L mode discharges is deduced from direct measurements of the turbulent Reynolds stress. The deviation increases with heating power. The turbulent poloidal viscosity is synthesized from fluctuation data, and is found to be comparable to the turbulent particle diffusivity. The intrinsic poloidal torque is deduced from synthesis, for the first time. PDFs of particle flux and Reynolds stress are obtained. Both exhibit fat tails and large kurtosis, suggesting that the momentum transport process represented by the Reynolds stress is not well described by quasilinear calculations.

1. MOTIVATION

Plasma poloidal mass flow and ExB flow are of great interest for their contribution to shearing and to the trigger mechanism for edge and core transport barriers. Experimental results on the DIII-D tokamak demonstrate a spin-up of main ion poloidal rotation at the transition from L mode to H mode. A transient increase in sheared poloidal flows and turbulent stress plays a critical role in triggering the L-H transition. Indeed, the poloidal rotation term is at least as important as the pressure gradient term in the radial force balance which determines radial electric field $E_r = \nabla_r P_i/(Z_i e n_i) - v_{\theta i} B_\phi + v_{\phi i} B_\theta$. Poloidal flow also plays an important role in triggering the transition to improved ohmic confinement on the HT-6M tokamak. A large excursion in the poloidal rotation of carbon impurity ions relative to the neoclassical prediction was associated with internal transport barrier formation in TFTR reversed shear plasmas. Standard neoclassical models have been used to calculate the poloidal rotation velocity at the edge of HL-2A tokamak. A significant deviation of poloidal rotation from the neoclassical prediction is deduced. Similar results were observed in previous studies on other tokamaks.

In the paper, we review the theory of poloidal rotation generation, including turbulence effects. The total Reynolds stress is measured. Analysis of fluctuation data is used to decompose the total stress into a turbulent viscous flux, which damps rotation, and a residual stress. Both contributions induce a velocity shift from the neoclassical value. The residual stress, proportional to the gradients which drive the turbulence, i.e. $\nabla T$, $\nabla n$, etc., effectively converts the free energy source to a drive of poloidal rotation through the turbulence. Virtually all models of turbulence effects on flows are based on a quasi-linear type approach, which tacitly presumes near-Gaussian statistics for the turbulence and transport. We investigate this by direct studies of the probability distribution functions of the edge particle flux $\Gamma_n$ and Reynolds stress $\Pi_{r,\theta}$. Our studies indicate the presence of fat tails on the PDFs, and kurtosis of $PDF(\Gamma_n)$, $PDF(\Pi_{r,\theta})$ well in excess of 10. This suggests that momentum transport occurs by avalanching, not diffusion. We conclude with discussion of ongoing work and future directions.

2. BACKGROUND

The theory of turbulence effects on mean poloidal flow via turbulent flux of momentum--Reynolds stress--has been studied and widely validated in the fusion community, since it was first proposed by P. H. Diamond and Y. B. Kim in 1991. Poloidal flow can shift relatively to its neoclassical value if the Reynolds stress $\langle \tilde{v}_r \tilde{v}_\theta \rangle$ has a nonzero divergence, i.e. $\partial_r \langle \tilde{v}_r \tilde{v}_\theta \rangle \neq 0$. This normally requires inhomogeneous turbulence. The Reynolds stress can be expressed in the form [1]:

$$\langle \tilde{v}_r \tilde{v}_\theta \rangle = -\chi_\theta \frac{\partial \langle v_\theta \rangle}{\partial r} + v_r^{eff} \langle v_\theta \rangle + \Pi_{r\theta}^{Res}. \quad (1)$$

The first term on the right-hand-side (RHS) represents the diffusive stress due to turbulent momentum diffusion, i.e. turbulent viscous flux. $\chi_\theta$ is the turbulent viscosity for poloidal flow. The second term represents the radial convection of poloidal momentum, and the third term is the residual stress, which has no leading dependence on $\langle v_\theta \rangle$ or $\partial \langle v_\theta \rangle / \partial r$. As a consequence of wave-flow momentum exchange, the residual stress drives an off-diagonal turbulent momentum flux and its divergence defines an intrinsic poloidal torque [2]. The residual stress is a function of profiles of density and temperature which drive the turbulence, i.e. $\Pi_{r\theta}^{Res} = \Pi_{r\theta}^{Res}(\nabla T, \nabla n)$, and of the turbulence intensity. The physical process where gradients drive rotation via residual stress can be understood by considering the analogy with a car engine that burns fuel and converts thermal energy so liberated into kinetic energy of a rotating wheel. In this physical picture, heating power and $\nabla T$ drive the turbulence, leading to profile relaxation and the generation of flow via turbulent stresses ultimately [3]. The divergence of the Reynolds stress shifts the poloidal flow from the neoclassical value. This is consistent with momentum balance [4]. Considering the case of a stationary small amplitude turbuence population and stationary flow:

$$\mu_{ii}^{(neo)}(\langle v_\theta \rangle - \langle v_\theta \rangle_{neo}) = -\partial_r \langle \tilde{v}_r \tilde{v}_\theta \rangle. \quad (2)$$

where the neoclassical viscosity coefficient (i.e. the flow damping rate) is $\mu_{ii}^{(neo)} \equiv \frac{1}{\tau_{ii}} \frac{\langle B^2 \rangle}{B_\theta^2} \mu_{00}$, $\langle v_\theta \rangle$ denotes poloidal rotation, $\langle v_\theta \rangle_{neo}$ represents neoclassical poloidal rotation and $\langle \tilde{v}_r \tilde{v}_\theta \rangle$ is Reynolds stress. Hereafter, $v_r^{eff}$ is ignored as it is negligible, so equation (3) is deduced. Note that both viscous diffusive and residual stress contribute to the deviation of $\langle v_\theta \rangle$ from its neoclassical value. Turbulence intensity gradients enter via $\partial_r \chi_\theta$ and $\partial_r (\Pi_{r\theta}^{Res})$.

$$\langle v_\theta \rangle = \langle v_\theta \rangle_{neo} + \frac{1}{\mu_{ii}^{(neo)}} \partial_r \left( \chi_\theta \frac{\partial \langle v_\theta \rangle}{\partial r} \right) - \frac{1}{\mu_{ii}^{(neo)}} \partial_r (\Pi_{r\theta}^{Res}). \quad (3)$$

The remainder of this paper is organized as follows: In Sec. 3, we introduce the experimental set up, including the diagnostic and discharge characteristics. Sec. 4 reports the experimental measurements of Reynolds stress in Ohmic discharge and L mode for different ECRH heating powers. We use the fluctuation data at the plasma edge to calculate the relative deviation of poloidal rotation from the neoclassical prediction. The poloidal velocity fluctuations are taken as ExB flow fluctuations, which are measured. Analysis is used to decompose the total stress into a turbulent viscous flux which damps rotation, and a turbulent residual stress, which drives rotation. In Sec. 5, we look beyond local, quasi-Gaussian models and calculate the probability distribution function of Reynolds stress and particle flux. Significant kurtosis and skewness are observed. Sec. 6 presents a discussion and conclusion.

3. EXPERIMENTAL SET UP

To measure physical quantities related to turbulent generation of poloidal rotation, a specially designed Langmuir probe array was built and installed on the outer mid-plane of HL-2A tokamak, shown by Fig 1. The spatial separation of adjacent tips in poloidal direction is 6 mm, and the spatial separation of adjacent tips in radial direction is 2.5mm. They are far less than the turbulent correlation length and so make the measurement reliable. The separations can be denoted by $d_\theta = 6mm$, $d_r = 2.5mm$, respectively. By using this Langmuir probe array, we can simultaneously measure plasma density, electron temperature, plasma velocities, Reynolds stress, etc.

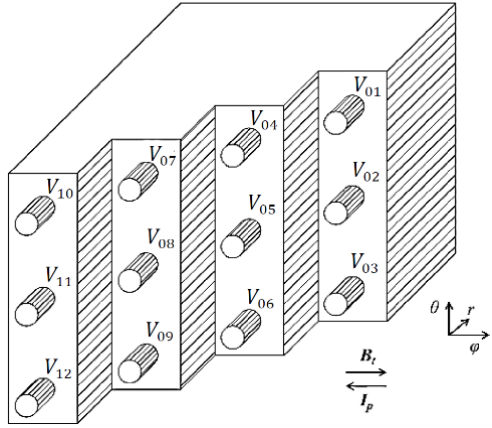

*FIG.1. A specially designed Langmuir probe array is installed on the outboard mid-plane of the HL-2A tokamak. Electron density and temperature, plasma potential, poloidal velocity, and Reynolds stress can be experimentally measured. $V_{10,+}$ and $V_{12,-}$ compose a double probe, combined with $V_{11,f}$ to form a triple probe. All the other channels are for floating potential measurement.*

Fluctuating ExB radial velocity is inferred by the measured potential difference in poloidal direction, $\tilde{v}_r = (\tilde{V}_{09,f} - \tilde{V}_{07,f})/2d_\theta B_t$. In a similar way, fluctuating ExB poloidal velocity is inferred by the measured potential difference in radial direction, $\tilde{v}_\theta = (\tilde{V}_{05,f} - \tilde{V}_{11,f})/2d_r B_t$. The Reynolds stress is computed as $RS = \langle \tilde{v}_r \tilde{v}_\theta \rangle$, where $\langle \cdot \rangle$ indicates the time average. Plasma density can be inferred from ion saturation current $I_{sat} = (V_{10,+} - V_{12,-})/R_s$, where $R_s = 25\Omega$ is the shunt resistor that the ion current flows through. Electron temperature is inferred as $T_e = (V_{12,-} - V_{11,f})/ln2$. Combining ion saturation current and electron temperature, we can infer plasma density as $n_e = I_{sat}/(0.61eA_{eff}C_s)$, where $C_s = \sqrt{kT_e/m_i}$ is the ion sound speed and $A_{eff}$ is the effective current collection area. The particle flux is computed as $PF = \langle \tilde{n} \tilde{v}_r \rangle$.

The experiments are conducted in Ohmic and ECRH heated L mode deuterium discharges, in a limiter configuration on the HL-2A tokamak. HL-2A is a medium-sized tokamak with a major radius of 1.65 m and a minor radius of 0.4 m. For a typical ohmic discharge on HL-2A when experiments are conducted, the plasma current is 160 kA and the toroidal magnetic field is roughly 1.35 Tesla. During time duration of 580-630 ms, the Langmuir probe was inside LCFS of the plasmas.

4. POLOIDAL ROTATION AND REYNOLDS STRESS

**4.1. Rotation and its deviation from neoclassical**

Neoclassical theoretical models for the calculation of poloidal rotation in tokamaks were proposed by Hazeltine, Hirshman-Sigmar, Kim-Diamond-Groebner [5] and Stacey-Sigmar. Here, the KDG model is used for calculation of the neoclassical poloidal rotation for the main ion "i". $v_{\theta i,neo}$ is given by:

$$v_{\theta i,neo} = \frac{B_\varphi K^i T_i L_{T_i}^{-1}}{Z_i e_i B^2}, (4)$$

where $B_\varphi$ denotes toroidal magnetic field, viscosity ratio $K^i \equiv \mu_{01}^i/\mu_{00}^i$, $T_i$ represents the main ion temperature, gradient scale length $L_{T_i}^{-1} \equiv -dlnT_i/dr$, $Z_i$ is proton number of the ion, $e_i$ is electric charge, and $B$ denotes magnetic field. The dimensionless collisionality $v_i^* \equiv v_{ii}qR/(v_{thi}\varepsilon^{3/2}) \sim 1$, shown in Fig. 2(a).

Thus, we use plateau regime results for estimation of the relevant parameters in the calculation of $v_{\theta i,neo}$. A significant deviation of the experimentally measured $E \times B$ poloidal velocity from the neoclassical prediction on both Ohmic (0 kW ECRH) and 300 & 700 kW ECRH heating power L mode discharges observed, as shown in Fig. 2(b).

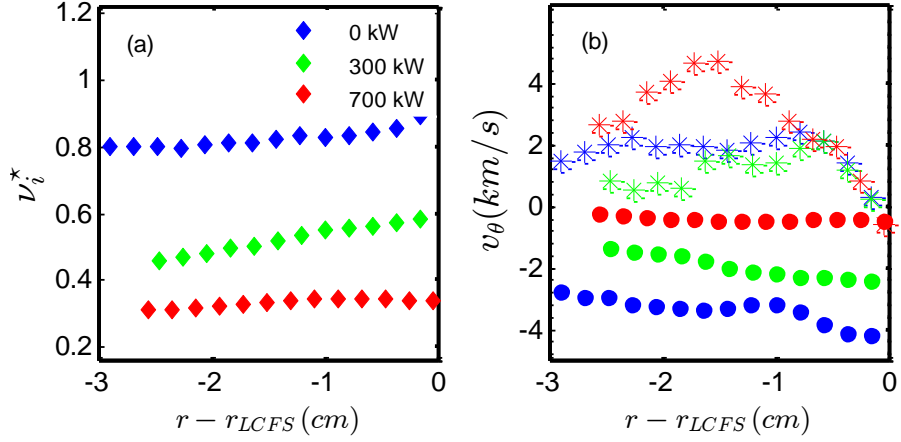

FIG.2. (a) Dimensionless collisionality $v_i^* \equiv v_{ii}qR/(v_{thi}\varepsilon^{3/2})$; (b) Neoclassical poloidal velocity and experimentally measured ExB flow velocity in Ohmic and L mode discharges. (Blue curve: Ohmic discharge; Green curve: 300 kW ECRH heating power; Red curve: 700kW ECRH heating power.)

The Reynolds stress is very likely the mechanism responsible for the deviation of poloidal velocity from neoclassical. Considering the case of stationary turbulence, $\langle v_\theta \rangle = \langle v_\theta \rangle_{neo} - \partial_r \langle \tilde{v}_r \tilde{v}_\theta \rangle / \mu_{ii}^{(neo)}$, where neoclassical viscous damping rate $\mu_{ii}^{(neo)} = \varepsilon^2 v_{thi}/qR$ for the plateau regime [6]. The relative deviation of poloidal rotation from the neoclassical prediction due to turbulent Reynolds stress and neoclassical viscosity is then given by $RD = -\frac{\partial_r \langle \tilde{v}_r \tilde{v}_\theta \rangle}{\mu_{ii}^{(neo)} \langle v_\theta \rangle_{neo}}$. In Fig. 3, we note that as heating power increases, the relative deviation of poloidal rotation from neoclassical increases significantly, and the slope of Reynolds stress inside LCFS increases. The experimental data demonstrates that the increased heating power leads to increased turbulence drive for the shear flow at the edge of the plasma. Studies of turbulent energy transfer are planned, so as to elucidate this process.

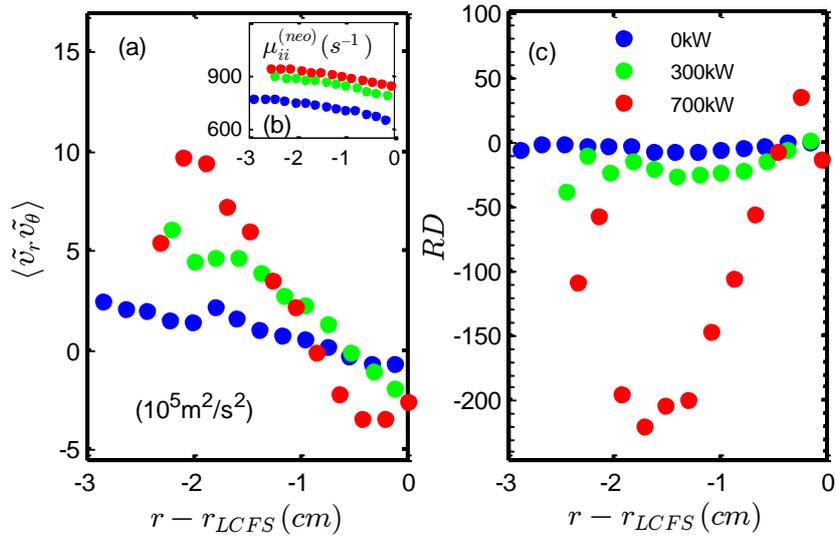

FIG.3. (a) Reynolds stress; (b) neoclassical viscosity $\mu_{ii}^{(neo)}$; (c) relative deviation of poloidal rotation from neoclassical RD

## 4.2. Decomposition of Reynolds stress

As we discuss in Sec. 2, the Reynolds stress can be decomposed to three terms. These are turbulent viscous flux, convection (eventually neglected) and residual stress, respectively. It's of natural interest to look for the contribution of these diffusive or non-diffusive stress to the turbulent generation of poloidal and ExB Flows.

The pinch velocity is assessed by gyro-kinetic calculations, i.e. $v_r^{eff} \cong \chi_\theta/R$. So the convection term contributes little to rotation generation when compared to viscous diffusion in the strong shear layer of the plasma edge. Thus, the deviation of the flow from neoclassical is due to turbulent viscous diffusion, which tends to relax the neoclassical gradient, and to $\partial_r(\Pi_{r\theta}^{Res})$. The latter converts free energy (i.e. $\nabla T, \nabla n$)—accessed by the turbulence—to intrinsic poloidal torque. Stationarity of the ExB flow and fluctuations can then be used to synthesize the turbulent viscosity from fluctuation data. Combined with the measurement of the Reynolds stress, the Residual stress can then be estimated [7].

The quasilinear expression for the ion flow diffusion coefficient (i.e. turbulent viscosity) is $\chi_\theta = \sum \frac{\langle \tilde{v}_r^2 \rangle |\gamma|}{(\omega - k_\theta v_\theta)^2 + |\gamma|^2}$, where the wave-particle decorrelation rate is $|\gamma| \sim \frac{1}{\tau_c}$, and $\tau_c$ is the decorrelation time. For modest turbulence, the spectral width exceeds the resonance width in $\chi_\theta$, so $\chi_\theta = \sum \langle \tilde{v}_r^2 \rangle \tau_{ac}$. Here, $\tau_{ac}$ is the spectral auto-correlation time. In general, $\tau_{ac} \leq \tau_c$, because $\tau_{ac}$ is spectrally integrated while $\tau_c$ is defined for each mode. Eddy-like structures exist in the edge shear layer. These on average persist for an auto-correlation time $\tau_{ac}$. The method for calculation of $\tau_{ac}$ and its dependence on range of fluctuation frequencies are shown in Fig. 4.

The turbulent momentum diffusion coefficient $\chi_\theta$ can be expressed in terms of the eddy radial velocity and the eddy auto-correlation time, via the relation $\chi_\theta = \langle \tilde{v}_r^2 \rangle \tau_{ac}$. The turbulent particle diffusion coefficient is $D_n$, which can be measured directly via $-\langle \tilde{n}\tilde{v}_r \rangle / \partial_r \langle n \rangle$. Fig. 5 shows a plot of these two diffusion coefficients. They are comparable and exhibit the same trends, as expected. This suggest that the synthesis of $\chi_\theta$ is plausible. Using the synthesized $\chi_\theta$, the measured ExB flow profile and the stationarity of the mean ExB flow, we have deduced the residual stress, following the approach of reference [7].

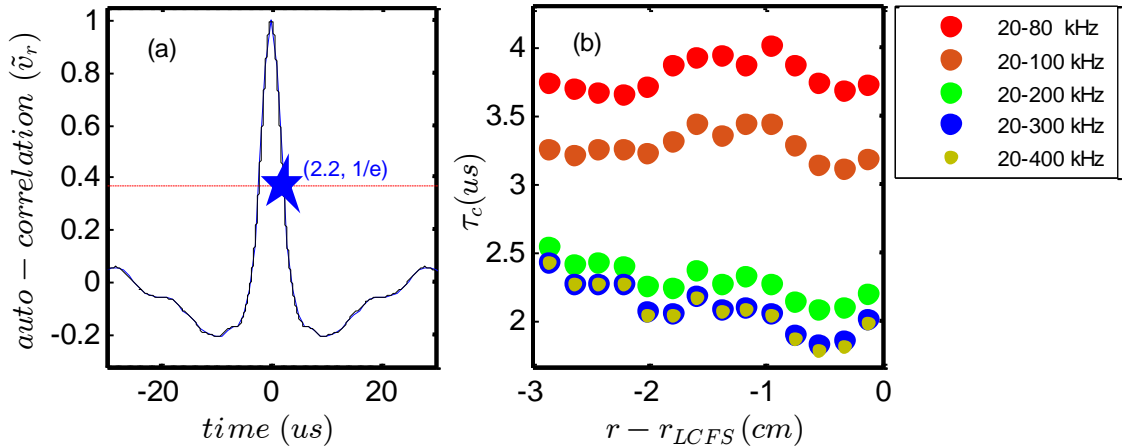

FIG.4. (a) Eddy auto-correlation time $\tau_{ac}$ is determined from the e-folding width of the auto-correlation function of $\tilde{v}_r$ fluctuations. (b) "Frequency saturation" in $\tau_{ac}$ shows the reason why we choose fluctuations with frequency $20 kHz < f < 300 kHz$.

The results are shown in Fig. 6. Fig. 6(h) shows the residual stress for three different powers. Note that the magnitude of $-\partial_r(\Pi_{r\theta}^{Res})$, i.e. the poloidal intrinsic torque, increases substantially at higher powers. Note too from Fig. 2(b) and Fig. 6(d), that while the neoclassical poloidal velocity is fairly flat, the total velocity develops a mean shear as power increases. This must be due to turbulence contributions. $-\partial_r(\Pi_{r\theta}^{Res})$ drives $\langle v_\theta \rangle$ away from neoclassical via its dependence on the strong intensity gradient.

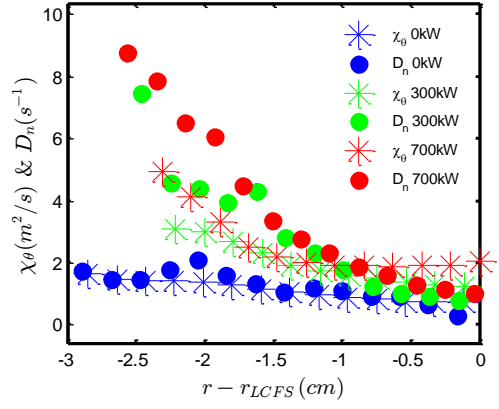

*FIG.5. Comparison of turbulent momentum viscosity $\chi_\theta$ to turbulent particle diffusivity $D_n$.*

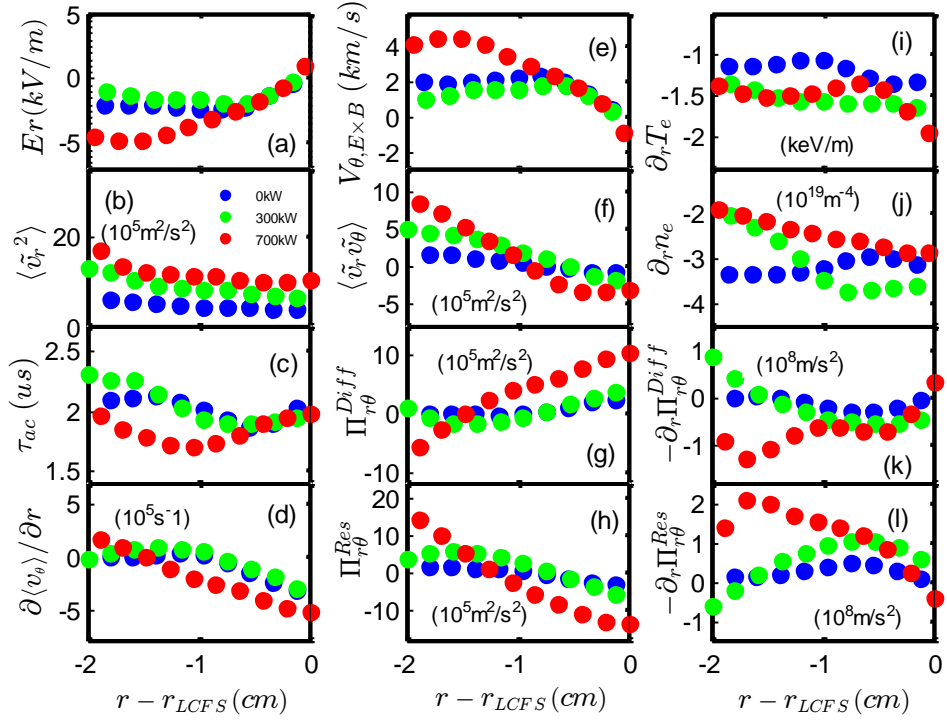

*FIG.6. Residual stress and its divergence. (a) radial electric field $E_r$; (b) mean square of ExB radial velocity fluctuations $\langle \tilde{v}_r^2 \rangle$; (c) auto-correlation time of $\tilde{v}_r$; (d) velocity shear $\partial \langle v_\theta \rangle / \partial r$; (e) ExB poloidal velocity; (f) Reynolds stress; (g) turbulent viscous flux; (h) residual stress; (i) local electron temperature gradient; (j) local electron density gradient; (k) turbulent viscous flux gradient; (l) residual stress gradient, i.e. the poloidal intrinsic torque.*

## 5. BEYOND THE QUASI-GAUSSIAN ANSATZ

Noting that virtually all models of turbulent momentum transport are based on quasi-gaussian, quasilinear models, we explore the statistics of the edge Reynolds stress and compare those to the statistics of the driving particle flux. The PDFs are shown in Fig. 7. Table 1 reports the skewness (S) and kurtosis (K) for $PDF(\Pi_{r,\theta})$ and $PDF(\Gamma_n)$. Note that $K_\Gamma \sim 15$ and $K_\Pi \sim 12$ are observed. These indicates that, both PDFs support significant "fat tails". This is not entirely surprising, since fluxes are a convolution of $\tilde{v}$, $\tilde{v}_\theta$ or $\tilde{n}$, and a cross phase. Thus, the Probability Distribution Functions of particle flux and Reynolds stress--both of which would be convolutions of quantities following quasi-Gaussian distributions--show pronounced deviations from a Gaussian distribution. Skewness and kurtosis exhibit sensitivities to heating power as well.

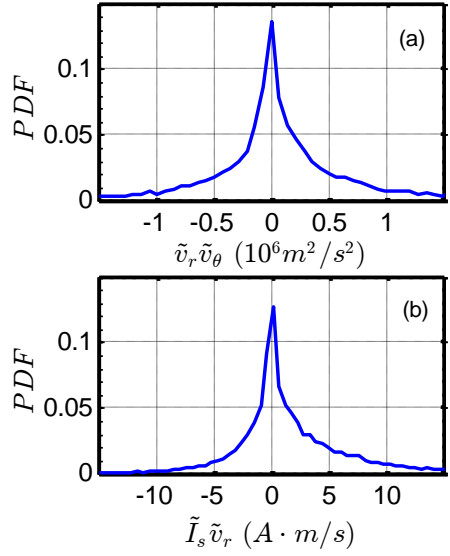

FIG.7. (a) Probability densities of Reynolds stress $PDF(\Pi_{r,\theta})$ and (b) of particle flux $PDF(\Gamma_n)$, at 1cm inside LCFS in an ohmic discharge. The turbulence frequency range is 20-300 kHz. The electron density fluctuation is replaced by ion saturation current fluctuation.

TABLE 1.  Skewness and kurtosis of PDF of Reynolds stress and particle flux at different radial position in an Ohmic discharge

| r-r$_{LCFS}$ (cm) | skewness | | kurtosis | |
| --- | --- | --- | --- | --- |
| | Reynolds stress | Particle flux | Reynolds stress | Particle flux |
| 0 | -1.2 | 1.5 | 13.9 | 18.3 |
| -1 | 0.3 | 2.2 | 12.2 | 15.0 |
| -2 | 0.3 | 2.2 | 10.0 | 14.0 |

Skewness and kurtosis of PDF of Reynolds stress and particle flux at different heating power at 1cm and 2cm inside LCFS are shown by Table 2 and Table 3, respectively.

TABLE 2.  Skewness and kurtosis of PDF of Reynolds stress and particle flux at different heating power at 1cm inside LCFS

| heating power (kW) | skewness | | kurtosis | |
| --- | --- | --- | --- | --- |
| | Reynolds stress | Particle flux | Reynolds stress | Particle flux |
| 0 | 0.3 | 2.2 | 12.2 | 15.0 |
| 300 | 0.1 | 2.0 | 15.7 | 13.6 |
| 700 | -0.3 | 1.9 | 11.5 | 13.9 |

TABLE 3.  Skewness and kurtosis of PDF of Reynolds stress and particle flux at different heating power at 2cm inside LCFS

| heating power (kW) | skewness | | kurtosis | |
| --- | --- | --- | --- | --- |
| | Reynolds stress | Particle flux | Reynolds stress | Particle flux |
| 0 | 0.3 | 2.2 | 10.0 | 14.0 |
| 300 | 0.7 | 2.3 | 9.6 | 13.5 |
| 700 | 1.2 | 2.6 | 11.1 | 17.0 |

These results—the first to study the $PDF(\langle \tilde{v}_r \tilde{v}_\theta \rangle)$—suggest that strongly non-Gaussian dynamics regulate poloidal momentum transport. One possibility is that avalanches of poloidal momentum regulate the transport. Further analysis is necessary, though. Computations and comparisons of the Hurst exponent for both fluxes are required and planned. These findings pose a significant challenge to the existing quasilinear paradigm for momentum transport.

6. DISCUSSION AND CONCLUSION

This paper explores aspects of the physics of turbulent transport of poloidal momentum, and its impact on the mean flow. The principle results reported are:

(a) The edge fluctuations, particle flux, Reynolds stress and mean ExB flow are characterized
(b) Significant deviation of mean poloidal flow from the neoclassical value is deduced. The deviation increases with heating power. Both diffusive and non-diffusive stresses contribute to the deviation.
(c) The turbulent poloidal viscous flux and residual stress are synthesized using fluctuation data. The turbulent poloidal viscosity is comparable to the turbulent particle diffusivity. The residual stress increases with heating power and exhibits a sharper gradient for higher powers.
(d) The PDFs of both Reynolds stress and particle flux exhibit fat tails and large kurtosis, suggesting non-Gaussian processes control momentum transport. It's significant that Reynolds stress has non-Gaussian features, despite the fact that momentum transport is a secondary process.

Future work will focus on:

(a) Using an independent measure of mean poloidal rotation to improve the synthesis.
(b) Using turbulence data to calculate $\Pi^{Res}$ from measured rotation.
(c) Calculating Hurst exponents for edge particle flux, Reynolds stress, and heat flux.
(d) Studies of the deviation from the neoclassical profile as edge heat flux increases, and approaches the L-H transition threshold.
(e) Experimental studies of Residual stress's parameter dependence and comparison with theoretical models.

ACKNOWLEDGEMENTS

We have benefitted from the First Chengdu Theory Festival 2018, where many of the relevant topics were discussed. We would also like to acknowledge useful discussions with Xuantong Ding and Bo Li.
This work is supported by the National Natural Science Foundation of China under Grant No. 11575055, 11705052 and 11875124, and by National Key R&D Program of China under 2017YFE0300405. The work is also supported by the U.S. Department of Energy, Office of Science, Office of Fusion Energy Sciences under Award Number DE-FG02-04ER54738.

REFERENCES

[1] Gürcan, Ö.D., et al., Intrinsic rotation and electric field shear, Phys. Plasmas **14** (2007) 042306.
[2] Diamond, P.H., et al., Physics of non-diffusive turbulent transport of momentum and the origins of spontaneous rotation in tokamaks. Nuclear Fusion **49** (2009) 571-576.
[3] Kosuga, Y., Diamond, P.H, and Gürcan, Ö.D., On the efficiency of intrinsic rotation generation in tokamaks. Phys. Plasmas, **17** (2010) 102313
[4] McDevitt, C.J., et al., Poloidal rotation and its relation to the potential vorticity flux, Phys. Plasmas **17** (2010) 042306.
[5] Kim, Y.B., Diamond P.H., and Groebner R.J., Neoclassical poloidal and toroidal rotation in tokamaks. Physics of Fluids B: Plasma Physics, **3** (1991) 2050-2060
[6] Diamond, P.H. and Kim, Y.B., Theory of mean poloidal flow generation by turbulence. Physics of Fluids B, **3** (1991) 1626-1633
[7] Yan, Z., et al., Intrinsic Rotation from a Residual Stress at the Boundary of a Cylindrical Laboratory Plasma, Physical Review Letters **104** (2010) 065002.